\documentclass[prl,nofootinbib,superscriptaddress]{revtex4}
\usepackage{graphicx}

\newcommand{\be}[1]{\begin{equation}\mbox{$\label{#1}$}}
\newcommand{\bea}[1]{\begin{eqnarray} \mbox{$\label{#1}$}}
\newcommand{\eea}{\end{eqnarray}}
\newcommand{\ea}{\end{eqnarray}}
\newcommand{\ee}{\end{equation}}
\newcommand{\Frac}[2]{\frac{\displaystyle #1}{\displaystyle #2}}
\newcommand{\ra}{\rangle}
\newcommand{\la}{\langle}

\def\npb#1#2#3{    {\it Nucl. Phys. }{\bf B #1} (#2) #3}

\def\pr#1#2#3{     {\it Phys. Rev. }{\bf   #1} (#2) #3}

\def\ijmpa#1#2#3{  {\it Int. J. Mod. Phys. }{\bf A #1} (#2) #3}
\def\zpc#1#2#3{    {\it Z. Phys. }{\bf C #1} (#2) #3}

\begin{document}
\hspace*{\stretch{1}}UB-ECM-PF-05/26\\
\hspace*{\stretch{1}} January 2006

\title{Vector meson decays 
from the Extended Chiral Quark Model}

\author{G. D'Ambrosio\footnote{E-mail: gdambros@na.infn.it}}
\affiliation{Istituto Nazionale di Fisica Nucleare, Sezione di Napoli \\
and Dipartimento di 
Scienze Fisiche,  Universit\`a di Napoli, I-80126 Naples, Italy.}
\author{D. Espriu\footnote{E-mail: espriu@ecm.ub.es}}
\affiliation{Departament d'Estructura i 
Constituents de la Mat\`eria and CER for Astrophysics,
Particle Physics and Cosmology,
 Universitat de Barcelona, 
Diagonal 647, 08028 Barcelona, Spain,}


\begin{abstract} 
We derive the the effective lagrangian that describes 
the interactions among vector, axial-vector mesons and pseudoscalars starting 
from the extended chiral quark model (ECQM). 
The results for the low-energy constants of this effective lagrangian
have a parametric resemblance with existing
predictions based on the Nambu-Jona-Lasinio model (except for some overall 
signs that we correct), but are numerically different. 
Therefore a precise measurement of these decay constants can shed some light
on the way chiral symmetry breaking is modelled in QCD.
Although most of the constants are poorly measured, 
comparison with phenomenology allows us to determine one of the parameters
of the ECQM that could not be fully determined in previous analyses. 
\end{abstract}


\maketitle

\section{Introduction}
The chiral quark model (CQM) \cite{1,2,Polosa:2000ym} 
 gives a good phenomenogical description of chiral symmetry breaking
and reasonable values for the Gasser Leutwyler cofficients, but does 
not describe  meson  states with masses $\sim 1 $ GeV
and does not provide a model for chiral symmetry breaking; it simply assumes 
that this takes place and incorporates the lowest 
dimensional operators compatible with the symmetry breaking pattern.
The  Nambu Jona Lasinio Model (NJL) \cite{3,7,brz,Pallante:1993jg} does provide a specific
model for chiral symmetry breaking by assuming strong attractive forces
in the scalar channel. It predicts a light narrow scalar partice, the elusive
$\sigma$ particle, the would-be chiral partner of the pion. But unitarization
studies combined with the large $N_c$ limit\cite{pelaez} suggest that such a 
particle is a dynamical resonance and not 
a truly QCD narrow resonance. Thus this simple model of chiral symmetry 
breaking is clearly disfavoured.

The possibility that the phenomenology of low energy QCD can be captured
by an hybrid model, where some features of both models are retained, was
 investigated in \cite{aet,ea}. The aim in these works was 
to write a very general low-energy model 
of QCD containing
all possible operators compatible with the symmetries of the model and then let
 phenomenology decide the respective importance of the different terms. 
The model is understood to be valid in the chirally broken phase 
(so like in the
CQM, no specific model of chiral symmetry breaking is assumed).
In this model the pion stands alone, and the partner of the $\sigma$ particle
(that is identified with a well established resonance, the $f_0$(980) in the
isoscalar channel) is the $\pi^\prime$(1300) in the isovector channel.
The authors named this model Extended Chiral Quark Model (ECQM).

In this work we shall explore some of the phenomenological consequences 
of the ECQM in the realm of vector and axial-vector decays. We shall argue
 later what is the phenomenological interest of understanding these decays. 
For us they
are basically a testing ground of the ECQM. 
It will be of interest to us also to
compare the predictions of the ECQM to those of the NJL model.
As we shall see the comparison is interesting to understand the 
criticality of the models on various parameters.

We shall first review the extended chiral quark model of \cite{aet,ea}.
After  we will present our derivation of the effective lagrangian for 
vector and 
axial-vector mesons, then present our numerical predictions for the low-energy
constants and conclusions.

\section{The effective chiral quark model}

The extended chiral quark model, ECQM, was introduced in \cite{aet,ea}. 
The reader 
is referred to these works for further details as we here present a succint 
description only. In Euclidean
conventions its lagrangian consists of three
different terms
\be{ECQM}
{\cal L}_{ECQM} = {\cal L}_{ch} + {\cal L}_{\cal M} + {\cal L}_{vec},
\ee
where
\bea{ECQM1}
 {\cal L}_{ch} &=&  {\cal L}_0 +
i\bar Q \left( \not\!\! D
 +   M_0 \right) Q
+ i \frac{4 \delta f_0}{\Lambda^2} \bar{Q} a_\mu  a_\mu Q\cr
&&+\frac{G_{S0}}{4N_{c} \Lambda^2}\
(\bar{Q}_L Q_R  +
\bar{Q}_R Q_L)^2 - \frac{G_{P1}}{4N_{c} \Lambda^2}
( -  \bar{Q}_L \vec\tau Q_R
+  \bar{Q}_R  \vec\tau Q_L)^2\cr
&&+\frac{G_{S1}}{4N_{c} \Lambda^2}
(\bar{Q}_L\vec\tau Q_R  +
\bar{Q}_R\vec\tau Q_L)^2 - \frac{G_{P0}}{4N_{c} \Lambda^2}
( -  \bar{Q}_L  Q_R
+  \bar{Q}_R  Q_L)^2,
\ea
\bea{massi}
 {\cal L}_{\cal M} &=&
i (\frac12 + \epsilon) \left(\bar Q_R {\cal M}
 Q_L + \bar Q_L  {\cal M}^\dagger  Q_R \right) \cr
&& + i (\frac12 - \epsilon) \left( \bar Q_R {\cal M}^\dagger  Q_L
+  \bar Q_L  {\cal M} Q_R\right)\cr
&& +  \langle  c_0\left({\cal M}  + {\cal M}^\dagger\right) 
 + c_5 ({\cal M} +{\cal M}^\dagger)a_\mu a_\mu
 + c_8 \left({\cal M}^2 + \left({\cal M}^\dagger\right)^2\right)\rangle  ,
\ea
and
\bea{vect}
{\cal L}_{vec} =&-& \frac{G_{V1}}{4N_c \Lambda^2} \bar Q \vec\tau \gamma_\mu
Q \bar Q \vec\tau \gamma_\mu Q -
\frac{G_{A1}}{4N_c \Lambda^2} \bar Q \vec\tau \gamma_5\gamma_\mu Q
\bar Q \vec\tau \gamma_5\gamma_\mu Q \cr
&-&  \frac{G_{V0}}{4N_c \Lambda^2} \bar Q \gamma_\mu Q
\bar Q  \gamma_\mu Q -
\frac{G_{A0}}{4N_c \Lambda^2} \bar Q \gamma_5\gamma_\mu Q
\bar Q  \gamma_5\gamma_\mu Q\cr
&+& c_{10} \langle U \bar L_{\mu\nu} U^\dagger \bar R_{\mu\nu}\rangle.
\ea
The notation we have used is the following: 
$Q$ are the quark fields written
in the 'constituent' or 'rotated' basis,
\be{constituent}
Q_L=u q_L,\qquad Q_R=u^\dagger q_R,\qquad u^2=U=\exp(2i\pi/F_0),
\qquad {\cal M}= u^\dagger m u^\dagger \ee
$m$ is the quark mass matrix,  $\not\!\!D$ is the covariant derivative
defined as
\be{va}
\not\!\! D \equiv \not\!\partial  + \not\! v - \gamma_5 \tilde g_A 
\not\! a, 
\ee
with the (antihermitian) fields
\be{vector}
 v_\mu = \frac12 \left( u^\dagger \partial_\mu u -
\partial_\mu u  u^\dagger 
+  u^\dagger \bar V_\mu u +
u \bar V_\mu u^\dagger - u^\dagger \bar A_\mu u +
u \bar A_\mu u^\dagger
\right)
\ee
\be{axial}
a_\mu = \frac12
\left( - u^\dagger \partial_\mu u -
\partial_\mu u u^\dagger 
-  u^\dagger \bar V_\mu u +
u \bar V_\mu u^\dagger + u^\dagger \bar A_\mu u +
u \bar A_\mu u^\dagger
\right),
\ee
where $\bar A$ and $\bar V$ are external axial and vector 
fields. 

The parameter $M_0$ is the so called `constituent' mass. $G_{S0}$,
$G_{P1}$, $G_{V1}$ and $G_{A1}$ are constants parametrizing the four-fermion 
interactions (indices denote the corresponding $J,I$ channels). These 
couplings will eventually be reduced and fixed by comparing with the 
physical values of vector meson masses. 
$\Lambda$ is a physical UV cut-off identified 
with the scale of
chiral symmetry breaking ($\simeq 1.4$ GeV).

The reader has by now undoubtedly noticed that ${\cal L}_{ch}$
contains the usual term operators in the CQM plus some four-quark operators
(reminiscent of the NJL). However we intend to describe physics 
in the chirally broken phase
and our fields include the pion matrix $u$, as befits an effective theory 
that should retain only the light degrees of freedom. Also, unlike in NJL 
the quark degrees of freedom appearing in (\ref{ECQM}) are from the 
very beginning 'constituent' 
quarks, quarks dressed by pions below the chiral symmetry breaking scale. 
The scalar and pseudoscalar four-quark couplings in (\ref{ECQM1}) 
need not be equal in order 
to preserve chiral symmetry (again unlike in NJL models).
 
An additional 
operator is allowed
by symmetry: ${\cal L}_{\cal M}$ contains the dependence on
current quark masses. Again, because of the possibility of 
including the $u$ field, 
the structure of this term is quite rich. Finally, ${\cal L}_{vec}$ includes
four quark operators in the vector and axial-vector channels. 

The term
\be{bare}
{\cal L}_0=-\frac{f_0^2}{4} \langle a_\mu a^\mu\rangle.
\ee 
as well as the operators whose coefficient are
$c_0$, $c_5$, $c_8$  and $c_{10}$ contain contributions from those
degrees of freedom with masses $\le \Lambda\simeq 1.4$ GeV. These
 ($c_0$, $c_5$, $c_8$  and $c_{10}$) 
contributions are typically small, the bulk of the contribution coming
from the light resonances. They are  
unimportant for our present discussion as is $\delta f_0$ in (\ref{ECQM1}).
 
The effective lagrangian in eq. (\ref{ECQM1}) 
is the most general\footnote{except for 
the fact that for simplicity not all isospin channels are included}
one compatible with the
principles of gauge and chiral invariance, $CP$ invariance and locality
that one can build out of quarks and pions up to, and including, operators
of dimension six. It contains four-fermion pieces somewhat reminiscent of
NJL, but the philosophy is different here: these terms typically will not have
large coefficients to trigger chiral symmetry breaking. No specific mechanism
is assumed for the latter, we just write an effective lagrangian that is
compatible with it. The vector field $\bar V$ contains a piece that commutes 
with $u$ describing the residual gluon interactions that ultimately ensure
confinement.  

Some of the constants and terms are somewhat non-standard. For instance, 
the na\"{i}ve QCD value for
the parameter $\epsilon$ is $\epsilon= 0.5$, but its actual 
value in the low energy theory is largely unconstrained. We shall return 
to this later.

After introducing auxiliary fields in all four channels, the effective 
lagrangian (\ref{ECQM}) becomes bilinear in the quark fields. The four-fermion
interaction is replaced by
\be{HS}
 \bar{Q}\left[i\widetilde\Sigma  - \gamma_5 \widetilde\Pi
 + \frac12 \gamma_\mu \widetilde V_\mu + 
\frac12 \gamma_\mu \gamma_5 \widetilde A_\mu\right] Q
+ 2 N_{c} \Lambda^2\left[\frac{\widetilde\Sigma ^2}{G_{S0}} +
\frac{(\widetilde\Pi^a)^2}{G_{P1}} + \frac{
\left(\widetilde V^{a}_\mu\right)^2}{4G_{V1}}
+ \frac{\left(\widetilde A^{a}_\mu\right)^2}{4G_{A1}}\right] 
\ee
and we include an integration over the
real auxiliary variables $\widetilde\Sigma,
 \widetilde\Pi^a, \widetilde V^{a}_\mu, \widetilde A^{a}_\mu$, defined by
$\widetilde\Pi \equiv \widetilde\Pi^a \tau^a/\sqrt{2}$,
$\widetilde V_\mu =  \widetilde V^{a}_\mu \tau^a/\sqrt{2}$, etc. (note that
the fields $\widetilde V^{a}_\mu$ and $\widetilde A^{a}_\mu$ are hermitian). 
This operation amounts 
to the replacement 
\be{replace}
v_\mu \to  {\cal V}_\mu =v_\mu - \frac12 i\widetilde V_\mu,\qquad
\tilde g_A a_\mu \to {\cal A}_\mu = 
\tilde g_A a_\mu -  \frac12 i\widetilde A_\mu,
\ee
and to the addition of scalar ($\Sigma$)  and pseudoscalar ($\Pi$) 
fields in the Dirac operator 
\be{dirac}
\Sigma =M_0 + \widetilde\Sigma +  \frac12 
\left({\cal M} + {\cal M}^\dagger \right)
+\frac{4 \delta f_0}{\Lambda^2} a_\mu a_\mu, \qquad
\Pi =  \widetilde\Pi + i \epsilon
\left({\cal M}^\dagger - {\cal M} \right), 
\ee
which becomes
\be{extdirac}
\hat D=\not\!\!\partial +{\cal \not\!\! V}-\gamma_5\tilde g_A
{\cal \not\!\! A} +
\Sigma + i\gamma_5 \Pi 
\ee

We can now integrate out the bilinear quarks and solve for the mass gap.
In the weak coupling regime the solution becomes 
\be{chibreaking}
\Sigma_0\simeq M_0 + m,
\ee
$m$ being the current quark mass. In practice the
constituent mass is large enough so that a derivative expansion in
inverse powers of $\Sigma_0$ makes sense at least for some
range of energies. We can thus write the
full quark-loop effective action. Retaining only the logarithmically enhanced
part we get \cite{aet,ea}
\bea{log}
{\cal L}_{1-loop} 
 &\simeq& \frac{N_c}{16\pi^2}  \ln\frac{\Lambda^2}{\Sigma_0^2}\,
 \langle
 (\Sigma^2 + \Pi^2)^2 + (\partial_\mu \Sigma)^2 + [D^{\cal V}_\mu, \Pi]^2\cr
&&- 4 ({\cal A}_\mu)^2 \Sigma^2 - \{{\cal A}_\mu, \Pi\}^2 
- 4i [D^{\cal V}_\mu, \Pi] \ {\cal A}_\mu \ \Sigma\ +
2i \partial_\mu \Sigma \{{\cal A}_\mu, \Pi\} \cr
&&- \frac16 \left( (F_{\mu\nu}^L)^2 + (F_{\mu\nu}^R)^2\right)\rangle.
\ea
$F_{L,R}$ are field strengths constructed with  ${\cal V}\pm {\cal A}$ and
$D^{\cal V}$ is the covariant derivative associated to the connection 
${\cal V}_\mu$.
 
In addition, we have the mass terms for the fields $\widetilde\Sigma$,
$\widetilde\Pi$, $\widetilde V_\mu$ and $\widetilde A_\mu$ coming from 
(\ref{HS}). In the axial channel there is some mixing between $a_\mu$
and $\tilde A_\mu$; the corresponding mass term reads
\be{diag}
\frac{N_{c} I_0 \Sigma_0^2}{4}
\langle \frac{1}{\bar G_{A}}
\widetilde A_\mu^2
+
\left(i 2 \tilde g_A a_\mu +   \widetilde A_\mu\right)^2\rangle.
\ee
The coupling $\bar G_A$ is introduced so as to give a natural scale for
the four-fermion terms (they turn out to be $\sim 0.1$)
\be{GA}  
\bar G_A=2G_{A1} I_0\frac{\Sigma_0^2}{\Lambda^2}\qquad
I_0=\frac{1}{4\pi^2}\ln\frac{\Lambda^2}{\Sigma_0^2}.
\ee
This mass term can be diagonalized  by defining
\be{diag2}
i 2 \tilde g_A a_\mu +   \widetilde A_\mu = 
i 2   g_A a_\mu +   \frac{1}{\lambda_{-}} A_\mu, 
\ee
with
\be{gadef}
 g_A = \frac{\tilde g_A}{1 + \bar G_{A}},
\ee
We refer to \cite{ea} for details.
$A_\mu$ is, finally the physical axial-vector field. In the vector field
there is no mixing. Of course 
we have to bear in mind that we are still in Euclidean space-time.
These expressions differ from the related expression in the extended NJL model
due to presence of a bare constant $\tilde g_A$. The constant
$\lambda_-$ is determined by requiring proper normalization
of the kinetic term for the $A_\mu$ field. One proceeds likewise for 
$\widetilde V_\mu$ finding that the properly normalized fields is
$V_\mu =\lambda_+ \widetilde V_\mu$. Furthermore one finds that
\be{normal}
\lambda^2_{+} =\lambda^2_{-} = \frac{N_c I_0}{6}. 
\ee

The values of the physical masses of the axial and vector mesons 
in terms of the parameters of the model can be found in Ref. \cite{ea}.
Ref. \cite{ea} concentrated on the implications of the model in two-point 
correlators. There it was seen that, after implementing the short distance
constraints coming from QCD via the Operator Product Expansion, in spite
of its relatively large number of parameters, the model could be well
constrained and some clear predictions emerged, comparing very favourably with 
the data. All the parameters in the ECQM can be thus determined (with one 
exception to be mentioned below). 

\begin{table}
\caption{The parameters of the ECQM as determined in reference \cite{ea}}
\label{tab:parameters}
\begin{center}
\begin{tabular}{|c||c|}
\hline
$\Lambda$ &  1.3 GeV \\
\hline
$\Sigma_0$ & 200 MeV  \\
\hline
$g_A$ & 0.55 \\
\hline
$\epsilon$ & 0.05 / -0.51  \\
\hline
\end{tabular}
\end{center}
\end{table}

There are two possible values for $\epsilon$ that are compatible 
with the fit of the two-point correlators and their subsequent matching 
to the OPE. This ambiguity will be resolved in this work. 

It was also seen in \cite{aet,ea} that the description of the 
low energy phenomenology that
the extended chiral quark model provides is clearly superior to that 
of the NJL model.

\section{Vector and axial-vector phenomenological lagrangians}

In what follows we want to explore other phenomenological consequences
of the extended chiral quark model by deriving the effective lagrangian
relevant for the decay of vector and axial vector mesons. All predictions will
be essentially parameter-free, as the model is rigidly fixed from
the two-point correlators as we have just indicated. The predictions
for vector meson decays at order $p^ 3$ are actually contained in 
the first term in the expansion of the determinant of the deneralized 
Dirac operator (\ref{log}). 
 
Let us introduce some notations and relations
\be{defin}
\nabla_{\mu} \, X \,  \equiv  \, \partial_{\mu} \, X \, + \, 
\left[ v_{\mu} \, , \, X \, \right],\qquad
X_{\mu\nu}\equiv \nabla_\mu X_\nu -\nabla_\nu X_\mu,
\ee
where $X=V,A$.
\be{defin2}
v^{\mu\nu}\equiv
\partial ^{\mu} \, v ^{\nu} \, -\partial ^{\nu} \, v^{\mu} \, +
\left[ v^{\mu} \, , \, v ^{\nu} \, \right]
\ee
\be{defin22}
 -\Frac{i}{2}  f_{+}^{\mu \nu}\equiv
v^{\mu\nu}- \Frac{1}{4}[ \, u^{\mu} \, , \, u^{\nu} \, ] 
\ee
\be{defin3}
f_{-}^{\mu \nu}\equiv 
\nabla ^{\mu} u^{\nu} - \nabla ^{\nu} u^{\mu}
\ee
where $u_\mu= -2i a_\mu$ is introduced to conform to the standard notation. 

Let us now consider the 
most general strong lagrangian linear in the vector field and up to  
${\cal O}(p^3)$ assuming
nonet symmetry. It reads \cite{EGL89,PradesZ}
\begin{eqnarray}
{\cal L}_V \, & = & \, - \, \Frac{f_V}{2 \sqrt{2}} \la \, V_{\mu \nu} \, 
f_{+}^{\mu \nu} \, \ra \, - \, \Frac{i g_V}{2 \sqrt{2}} \, \la 
V_{\mu \nu} \, [ \, u^{\mu} \, , \, u^{\nu} \, ] \ra  \nonumber \\
& &  \, + \, i \alpha_V \, \la \, V_{\mu} \, [ \, u_{\nu} \, , \, 
f_{-}^{\mu \nu} \, ] \, \ra 
 \, + \, \beta_V \, \la \, V_{\mu} \, 
[ \, u^{\mu} \, , \, \chi_{-} \, ] \, \ra 
\end{eqnarray}
In the above expresion
\be{massterm}
\chi_\pm=2B_0(u^+ m u^+ \pm u m ^\dagger u)\qquad
B_0(1 {\rm GeV})\simeq 1.5\  {\rm GeV}
\ee
We do not include the odd-parity part in the above lagrangian (proportional
to $\epsilon^{\alpha\beta\mu\nu}$).
For axial-vector fields
\begin{eqnarray}
{\cal L}_A \, & = & \, - \, \Frac{f_A}{2 \sqrt{2}} \, \la \, A_{\mu \nu} \, 
f_{-}^{\mu \nu} \, \ra \, + \, i \alpha_A \, \la \, A_{\mu} \, [ \, 
u_{\nu} \, , \, f_{+}^{\mu \nu} \, ] \, \ra 
 \, + \, \gamma_1 \, \la \, A_{\mu} \, u_{\nu} \, u^{\mu} \, u^{\nu} \, 
\ra \, \nonumber \\
& & \, + \, \gamma_2 \, \la \, A_{\mu} \, \{ \, u^{\mu} \, , \, u^{\nu} \, 
u_{\nu} \, \} \, \ra 
\, + \, \gamma_3 \, \la \, A_{\mu} \, u_{\nu} \, \ra \, \la \, 
u^{\mu} \, u^{\nu} \, \ra \, + \, \gamma_4 \, \la \, A_{\mu} \, u^{\mu} \, 
\ra \, \la \, u_{\nu} \, u^{\nu} \, \ra 
\end{eqnarray}
Again the terms containing $\varepsilon_{\mu \nu \rho \sigma}$ will not be 
considered in this work. Note that there is some ambiguity in the choice 
of overall signs. We choose eventually the sign of the axial field so as
to conform to the usual conventions (implying a positive $f_A$).

In order to make contact with phenomenology, we have to Wick rotate the
Euclidean effective lagrangian we have obtained.
Using the previous expressions, 
from the extended Chiral Quark model the following predictions emerge
in the vector and axial-vector meson sector
\be{result}
f_V^2=N_cI_0/6,\quad\quad  g_V=f_V\Frac{1-g_A^2}{2},\quad\quad
\alpha _V=f_V\Frac{g_A^2}{2\sqrt{2}},\quad\quad
\beta _V=f_V \Frac{3g_A M_0 \epsilon}{2\sqrt{2} B_0},
\ee
\be{result2}
f_A=f_V g_A,\quad\quad \alpha _A=f_V\Frac{g_A}{2\sqrt{2}},\quad\quad
\gamma _A ^1 =-f_V \Frac{g_A(1-g_A^2)}{2\sqrt{2}},\quad\quad
\gamma _A ^2 =f_V \Frac{g_A(1-g_A^2)}{4\sqrt{2}}.
\ee
These are the predictions of the ECQM. 

When comparing to the predictions of the NJL model \cite{PradesZ},
 we note that,
although the details of the expressions between our results and those
of the NJL model obviously differ, when looking at the leading term in NJL
there is an overall change of sign in the axial-vector couplings ($\alpha_A,
\gamma_A^1,\gamma_A^2$) and also in
$\beta _V$ and $\alpha _V$, perhaps due 
to different conventions in Minkowski and Euclidean space. An overall 
change of sign everywhere is
 of course undetectable.

\section{Numerical analysis and conclusions}

The previous predictions, making use of the 'best fit' presented in
Table \ref{tab:parameters} lead
to the set of numerical values quoted in Tables 
\ref{tab:predV} and 
\ref{tab:predA}.

While there is no need to stress here the relevance of $g_V\ , f_V $ and 
$f_A$ it is worth emphasizing the phenomenological impact of the
other couplings.  Vector meson dominace in weak non-leptonic kaon 
decays have been studied
accurately
\cite{Ecker:1992de,D'Ambrosio:1997tb}.
In fact it has been shown in Ref.
\cite{D'Ambrosio:1997tb} that there are several cases where
the remaining couplings are particularly 
interesting due the  vanishing of the contributions due to  $g_V$ and $ f_V $
are: see $K\to 2\pi /3 \pi$ and  $K\to\pi ^+\pi ^0 \gamma   $
\cite{D'Ambrosio:1997tb}.
The couplings
in ${\cal L}_V$ and ${\cal L}_A$ can be determined, in principle, 
from the phenomenology
of the vector meson decays. $|f_V|$ and 
$|\alpha_V|$ could be obtained from the experimental widths
\cite{Eidelman:2004wy}
of $\rho^0 \rightarrow e^+ e^-$, $\omega \rightarrow \pi^0 \gamma$, 
$\omega \rightarrow \pi \pi \pi$ and $\rho \rightarrow \pi \pi \gamma$,
respectively, while $g_V$ and $\beta_V$ enter in $\rho \rightarrow 
\pi \pi$. 

As for the axial-vector couplings they can be determined from the decays
$a_1^+\rightarrow 3 \pi$  ($\gamma _1 $, $\gamma _2 $, $\gamma _3 $ and $\gamma _4  $) and 
$a_1^+\rightarrow  \pi^+ \gamma $  ($ f_A $ and $\alpha _A $)\cite{PradesZ}.
 However 
unfortunately data are not precise enough
to go beyond a good determination of $ f_A $.

In Table 2 we collect the experimental determinations (when
available). As we have emphasized the predictions are absolutely 
rigid, as all the free parameters in the model are fixed beforehand from the 
two-point functions.

\begin{table}

\caption{The predictions of the ECQM for the vector couplings 
compared to experiment}
\label{tab:predV}
\begin{tabular}{|c||c|c|}
\hline
& ECQM & Experiment\\
\hline
\hline
$f_V$ & input &     0.20 \\
\hline
$g_V$ &  0.07    & 0.09 \\
\hline
$\alpha_V$ & 0.02    & - \\
\hline
$\beta_V$ & -0.008 &-0.018 \\
\hline
\end{tabular}
\end{table}

\begin{table}
\caption{The predictions of the ECQM for the axial-vector couplings}
\label{tab:predA}
\begin{tabular}{|c|c|c|}
\hline & ECQM  & Experiment \\
\hline
\hline
$f_A$ &  0.11 &        0.097  \\
\hline
$\alpha_A$ & 0.04 &      - \\
\hline
$\gamma_1$ &  -0.03     &  - \\
\hline
$\gamma_2$ &  0.01       &- \\
\hline
$\gamma_{3,4}$ & ${\cal O}(1/\sqrt{N_c})$ 
&- \\
\hline

\end{tabular}

\end{table}

When comparing the experimental value for $\beta_V$ with the theoretical
prediction of the model, this favours the value $\epsilon=-0.5$. That solves
the ambiguity in the determination of $\epsilon$ we alluded to before and fixes
completely the leading coupling constants of the ECQM.

It is unfortunate that except for $g_V$ and $\beta_V$ the other couplings 
in this parity even sector are not measured yet. In some of them we get
results that clearly differ numerically from the predictions of the 
NJL model and therefore they provide
a clear test of the mechanisms of chiral symmetry breaking. Their measurement
is clearly interesting.

Even if we have reduced ourself to the study of 
non-anomalous vector and axial coupling
some interesting conclusions on NJL and ECQM  can be drawn.
Our numerical values certainly differ from the NJL ones and thus measuring the
low energy constants related to meson decays into pseudoscalars can 
be particularly
telling about the mechanisms of chiral symmetry breaking in QCD and its 
modelization. 
We have been able to resolve the ambiguity in the 
determination of the $\epsilon$ 
parameter in the ECQM.

Particularly useful are some VMD couplings  which could be measured 
in the near future and might be phenomenological relevant in K-meson decays.

\section*{Acknowledgements}

This work ~was partially supported by IHP-RTN,
EC contract No.\ HPRN-CT-2002-00311 (EURIDICE). G.D. thanks the warm hospitality 
of Universitat de Barcelona,
 D.E. would like
to acknowledge the support of the grant FPA2004-04582 and the hospitality 
of INFN-Naples. We would like to thank A.Andrianov for a long discussion 
concerning conventions and Wick rotation and J.Portoles and 
J.Prades for several comments.

\end{document}